\documentclass[twocolumn]{article}
\usepackage{graphicx}
\usepackage{amsfonts}
\usepackage{amssymb,amsmath,amsthm}
\usepackage{euscript}
\usepackage{color}
\newtheorem{definition}{Definition}

\usepackage{algorithm}
\usepackage{algpseudocode}
\usepackage{balance}
\usepackage{url}

\algblockdefx{Sync}{EndSync}[1]{\textbf{sync} #1}{\textbf{end sync}}

\begin{document}
\title{Preserving Co-Location Privacy in Geo-Social Networks}
\author{Matteo Camilli \\
Dipartimento di Informatica e Comunicazione\\
Universit\`a degli Studi di Milano, Italy\\ 
Email: matteo.camilli@unimi.it}

\maketitle

\begin{abstract}
The number of people on social networks has grown exponentially. Users share very large volumes of personal informations and content every days. This content could be tagged with geo-spatial and temporal coordinates that may be considered sensitive for some users. While there is clearly a demand for users to share this information with each other, there is also substantial demand for greater control over the conditions under which their information is shared. Content published in a geo-aware social networks (GeoSN) often involves multiple users and it is often accessible to multiple users, without the publisher being aware of the privacy preferences of those users. This makes difficult for GeoSN users to control which information about them is available and to whom it is available. Thus, the lack of means to protect users privacy scares people bothered about privacy issues. This paper addresses a particular privacy threats that occur in GeoSNs: the Co-location privacy threat. It concerns the availability of information about the presence of multiple users in a same locations at given times, against their will. The challenge addressed is that of supporting privacy while still enabling useful services.
\end{abstract}

\section{Introduction}
\label{sec:intro}
The great availability of social network services and mobile devices with internet connectivity and intograted GPS enable Geo-aware Social Networks (GeoSNs).
Several different existing GeoSNs allow users to share their location (frequently, their exact location on a map) and other types of information, but have extremely limited privacy settings. Typically, they only allow users to specify location informations with higher granularity (for example, a city), or a list of individuals with whom they would be willing to share their locations at any time \cite{tsai2010location}. While there is clearly a demand for users to share this information with each other, there is also substantial demand for greater control over the conditions under which their information is shared, and a number of recent papers demonstrate that individuals are concerned about privacy in this domain \cite{Consolvo05, Sadeh09, Tsai09}.


Thus, privacy in social networks is a hot topic, and reports indicate that an increasing number of users are concerned about privacy issue, enough to leave GeoSNs \cite{guardianArticle}. In GeoSNs, exact locations of users are published and can be red by multiple users. Thus, potentially untrusted entities may exploit these to infer sensitive information about the users and make some unwanted focused actions. Recent studies has been performed on different aspects of user privacy that are potentially at risk \cite{Freni10, Vicente11}. These works examine specific privacy threats and try to give possible solutions. In particular the \emph{Location Privacy}, the \emph{Absence Privacy} and the \emph{Co-Location Privacy} threats are araised. While the first two problems are widely discussed in \cite{Freni10}, the \emph{Co-Location Privacy} still remains unexplored, as far as we know.

In most GeoSNs, an adversary might be able to observe the presence of multiple users in the same place and some users consider such co-location to be sensitive. Thus, disclosing this information to an adversary constitutes a co-location privacy violation. None of the currently known GeoSNs supports co-location privacy \cite{Vicente11}, and specifying privacy preferences related to co-location can be challenging.

The objective of this paper is to explore this problem and provide preliminary techniques that enable users to specify their privacy preferences and then enforce these preferences. Our approach is based on the studies performed in \cite{Freni10}. We apply meta-data generalization in order to make not possible for any set of resources violate any users preference, also taking into account constraints on the maximum velocity of user movement. The proposed technique enforce the co-location privacy by computing appropriate spatial enlargement (where possible) in resource publication.

Although privacy has been studied extensively in location-based services and social networks \cite{Mascetti09, Kalnis07, Mascetti07, Zhong07}, we are not aware of any studies that consider co-location privacy in the GeoSN setting.

Finally, some existing GeoSN services offer some form of control of the geo-tags of resources, e.g., by enabling tags at coarse granularities such as the city level (e.g., Google Latitude), but much finer controls are necessary to avoid the privacy threats considered in this paper.
The contributions of the paper are the following:

\begin{itemize} 
  \item Formalization of co-location privacy threat in GeoSN and adversary attacks.
  \item Proposals of means of expressing privacy preferences.
  \item A privacy preserving technique that guarantees the enforcement of user preferences.
\end{itemize} 

The rest of the paper is organized as follows. Section \ref{sec:formal} formally characterizes the co-location privacy threat in GeoSNs, the adversary model and how users can specify their privacy preferences; Section \ref{sec:algo} describes the algorithm of the proposed GeoSN privacy preservation technique; Section \ref{sec:app} discuss the applicability of our technique to existing GeoSNs; Section \ref{sec:concl} reports our conclusion.

\section{Problem Formalization}
\label{sec:formal}
In this section, we formally describe the assumed GeoSN service resources and the privacy threats we address. Then we define how users can express their privacy preferences, the adversary model, and sufficient conditions for satisfying a user's privacy preferences.

\subsection{GeoSN Resources}
\label{sec:res}
A GeoSN service allows its users to publish a resource (e.g., a picture, a text message, a check-in) tagged with the current location and time, as well as a set of users related to the resource. A resource is either automatically tagged (e.g. an integrated GPS can provide location and time), or manually tagged. Since resources and their tags become available to other users as well as to service providers, we are concerned with the privacy violations that the publication can lead to.
Formally, a $resource~r$ is a tuple:

\begin{equation} \label{eq:res1}
 \langle U, T, S, C \rangle
\end{equation}

where the elements are meta-data tags, with $r.U$ being a set of identifiers of users, $r.S$ being a spatial tag, $r.T$ being a temporal tag and $r.C$ being the resource itself. In the following, when referring to a $resource~r$, we assume that all the users in $r.U$ are in the location $r.S$ at the time $r.T$. We denote the user that makes $r$ as $owner(r)$. Note that:

\begin{equation} \label{eq:users_tag}
\begin{split}
 r.U \supseteq owner(r) ~\wedge \\ \forall u \in r.U \setminus \{owner(r)\}, ~friend(owner(r),u)
\end{split}
\end{equation}

where $friend$ is a ``friendship" relation between users. Location of a resource could be recorded  at the finest available resolution (a point in the appropriate domain) or with higher granularity, that is a larger area. Time of a resource is a timestamp with date and time. In this paper we refer to a \emph{real time publication} model. This means that each resources has an accurate timestamp and users can't publish resources referring to the past. This model may include for example proximity services, micro-blogging, and social navigation services. We denote the set of resources of the GeoSN as $R \subseteq \mathcal{R}$ (the resources domain).

%
%

In accordance to a consolidated idea \cite{Freni10}, we based our approach on the \emph{generalization} of the resources before publication. In particular, when we identify a \emph{resource r} that violate the privacy of a user (or a set of users) we apply a function $g$ that takes a \emph{resource r} and generate a \emph{resource r'} that doesn't violate the privacy of any users. Formally, a \emph{generalization} function is $g: \mathcal{R} \longrightarrow  \mathcal{R} $:

\begin{equation} \label{eq:genfunc}
\begin{split}
  g(r=\langle U, T, S, C \rangle) = (r'=\langle U', T, S', C \rangle)
\end{split}
\end{equation}

where $U' \subseteq U$ and $S \subseteq S'$. The function $g(r)$ takes into account the privacy preferences expressed by the users and try to make a new resource that does not violate these preferences.

In the following we recall two basic concepts (first introduced in \cite{Freni10}) that will be used in the next sections. A resource $r$ is $reachable$ from another resource $r'$ if each spacial point in $r$ are reachable from some spacial points in $r'$, in the time interval $|r.T-r'.T|$ moving with an acceptable speed. Formally:

\begin{definition}[Reachability]
Given a velocity $V_{max}$ and two resources $r, r'$, we say that $r$ is $reachable$ from $r'$ if:
\begin{equation} \label{eq:reach}
\begin{split}
\forall s \in r.S, ~ \exists  s' \in r'.S :  \frac{d(s,s')}{|r.T-r'.T|}<V_{max}
 \end{split}
\end{equation}
where d(s, s') compute the distance between the two spacial points and $V_{max}$ is the maximum acceptable medium speed for a user.
\end{definition}

Two resources are $independent$ if they have not users in common or if their spatial distance is small when compared with the temporal distance. Formally:

\begin{definition}[Independence]
Given two resources $\langle r,r' \rangle \in R \times R$, we say that $r$ and $r'$ are independent (and we denote it with $r \perp r'$) if:
\begin{equation} \label{eq:indip}
\begin{split}
r.U \cap r'.U = \emptyset ~\vee \\
 (r ~reachable ~from ~r'  ~\wedge~ r' ~reachable ~from ~r)
 \end{split}
 \end{equation}
\end{definition}

\begin{figure}[ext]
\centering
 \includegraphics[width=0.6\columnwidth]{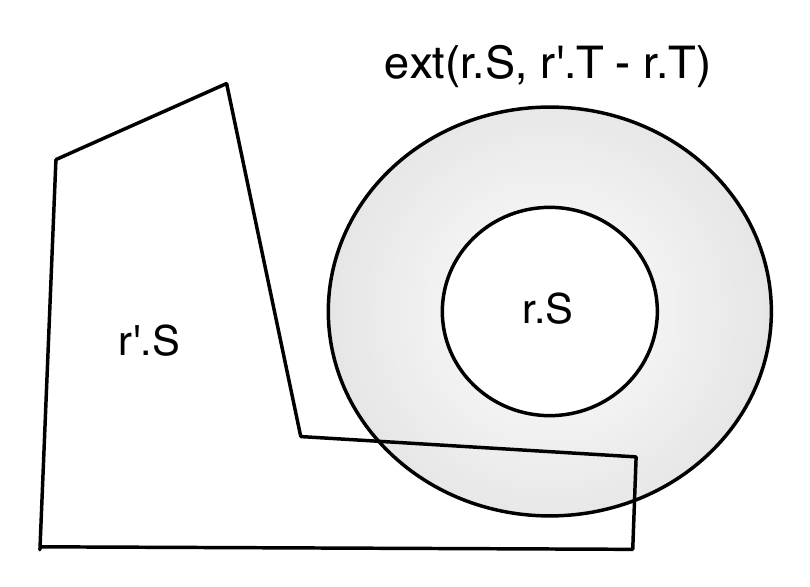}
\caption{Consistency of resources.} 
\label{fig:ext}
\end{figure}

$Independence$ of resources ensures that each resource doesn't restrict the informations given by any other resources. This property allows to avoid a particular attack from the adversary. For example, suppose that in Fig.~\ref{fig:ext} are represented the spatial informations of two consecutive resources that involve the user $u$. The $ext$ function computes the area in which any user located in $r.S$ at the time $r.T$, can be located at the time $r'.T$ (considering the maximum speed of users).
Thus, the adversary can infer that only a subset of $r'.S$ is a possible location for $u$. we will clarify the need of $independence$ of resources in the section \ref{sec:priv_pres}

\subsection{Co-Location Privacy Threat}
\label{sec:coloc}
We consider geographical location and temporal location as sensitive information that sometimes users want to maintain private. When an adversary can associates user's identity with these kind of informations without the user's consensus, a privacy violation has occurred. This paper focuses on the \emph{Co-Location privacy} problem.
In this section we give a description of this privacy threat and we give a description of how users could describe their privacy preferences in order to protect themselves from this problem.

A typical \emph{Co-Location privacy} threat example is a user that doesn't want to let people know that (s)he is located in a specific place at a specific time with her/his secret lover. For instance, imagine that Alice and Bob are secret lovers and they having a drink together in a pub. They don't want to let other people know their secret meeting, but Bob sees his friend Charlie that updates his status, writing ``just met Bob at a pub!" and tagging the post with 12:02 p.m., 24 West 35th Street. A bit later Alice sees her friend Juliette that updates her status, writing ``just met Alice at a pub!" and tagging the post with 12:10 p.m., 24 West 35th Street. A user with access to both posts (e.g., a friend of Bob and Alice) can infer that Alice and Bob are co-located without their consensus.

The disclosure of a co-location can occur in two different ways: a $direct$ way and an $indirect$ way. The $direct~disclosure$ occur when a single $resource ~r$ co-locates a set of users $U$ that don't want to be co-located. The $indirect$ $disclosure$ occur when a pair of $resource ~r,r'$ co-locate two different subset of $U$ in a ``small" area and with ``small" time difference (this is the case presented in the above example). We discuss what we mean with ``small" in section \ref{sec:priv_pres}.

This privacy threat can be addressed by offering the users means of controlling the location information to be disclosed. This means include different kinds of privacy preferences that users can express. We model the preferences for avoid co-location privacy threat for a user $u$ as a tuple $\varphi$:

\begin{equation} \label{eq:priv_pref}
 \langle E, A, T, D \rangle
\end{equation}

where $\varphi.E$ and $\varphi.A$ are sets of users, $\varphi.T$ is a time interval and $\varphi.D$ is a spatial distance. In particular,

\begin{itemize} 
  \item $\varphi.E$ is called $Excluding~Set$ and represent the set of users with whom $u$ doesn't want to be co-located.
  \item $\varphi.A$ is called $Adversary~Set$. Co-location of $u$ with any users in $\varphi.E$ is allowed if the co-location includes any users in $\varphi.A$ as well (we'll explain better this concept in the next section).
  \item $\varphi.T$ is the time interval in which $u$ doesn't want to be co-located. $\varphi.T$ have a starting time ($t_{start}$) and an ending time ($t_{end}$). $\varphi.T$ is bounded ($t_{end}-t_{start} \leq T_{max}$).
  \item $\varphi.D$ is the minimum distance within $u$ doesn't want to be co-located. $\varphi.D$ is bounded ($\varphi.D \leq D_{max}$).
\end{itemize}

For instance, a possible co-location privacy preference for Alice could be ``Don't reveal my co-location with Bob in less than 50 meters and during the evenings, unless Bob's wife (Mary) is there as well.". This privacy preference can be represented by a recurring (infinite) set of $\varphi$ tuple of the type: 
\\$\langle$ Bob, Mary, 11th July '11 (19:00 p.m.) - 11th July '11 (23:00 p.m.), 50 m. $\rangle$
\\$\langle$ Bob, Mary, 12th July '11 (19:00 p.m.) - 12th July '11 (23:00 p.m.), 50 m. $\rangle$
\\$\langle$ Bob, Mary, 13th July '11 (19:00 p.m.) - 13th July '11 (23:00 p.m.), 50 m. $\rangle$
\\ and so on...

starting from the day in which Alice has set this preference. We indicate the set of the privacy settings of the $user ~u$ with $\Phi(u)$.

Intuitively, when a user want to make a resource $r$ that violates this preference we apply a generalization function $g$ to $r$ that obfuscate the meta-data of $r$, in order to preserve the privacy of Alice and Bob. Obfuscation in our case means the enlargement of the area expressed by $r.S$ or the removal of some user in $r.U$, so that an adversary is not able to infer exact informations about the victims.

\subsection{Adversary Model}
\label{sec:advers}
The adversary is a user of the GeoSN that want to use published resources to infer sensitive informations about other users (victims). We assume that the adversary has access to all the resources published by all the users. This conservative approach is not a realistic context but it has two important effects: (a) if the co-location privacy is preserved against an adversary that has access to all the resources of the GeoSN, it is also preserved against an adversary that has a restricted access to the resources; (b) this approach permits to avoid any chance by an adversary to exploit future friendship relations (or friendship relations not in common with the victims) in order to gain access to additional sensitive informations about users. We also assume that the adversary knows the generalization technique used to generalize resources before publication, but (s)he doesn't know the privacy preferences of other users because these informations are only available if (s)he has access to other accounts (and we assume that (s)he doesn't have it).

When  a resource $r \in R$ states that a user $u$ is located in the area $r.S$ at the time $r.T$, the adversary can assume a uniform probability distribution of user location, that is:

\begin{equation} \label{eq:prob1}
 \forall ~ s \in r.S, ~ P(loc(u)=s)=p
\end{equation}

where $loc(u)$ indicates the exact spatial location of the user $u$. Instead $\forall t \neq r.T$, the adversary can assume that u can be located in a larger area:

\begin{equation} \label{eq:prob2}
\begin{cases} 
  P(loc(u) = s)>0, & \mbox{if } s \in ext(r.S, |t-r.T|) \\ 
  P(loc(u) = s)=0, & \mbox{if } s \notin ext(r.S, |t-r.T|)
\end{cases}
\end{equation}

with $ext$ being the function defined in section \ref{sec:res}.
This means that we consider not null the probability that $u$ is located in a larger area if $u$, moving with an acceptable medium speed, can be in this area after the time interval $|t-r.T|$.

Concerning the co-location privacy of a user $u$, we consider it as preserved if $\forall ~ \varphi \in \Phi(u)$, the adversary doesn't have any chance to consider null the probability that $u$ is at least $\varphi.D$ far from any users in $\varphi.E$ in the temporal interval $\varphi.T$.

\begin{definition}[Co-Location Privacy preservation of a user u]
The Co-Location Privacy of $u$ is preserved if:
\begin{equation} \label{eq:priv_pres_u}
\begin{split}
 \forall \varphi \in \Phi(u), ~ \forall ~ t \in \varphi.T, ~ \forall ~ e \in \varphi.E \\
  P(d(u, e)>\varphi.D)>0
  \end{split}
\end{equation}
\end{definition}

and this holds as long as doesn't exists any set of resources that makes null this probability.

\begin{definition}[Co-Location Privacy preservation]
A GeoSN preserve the Co-Location Privacy if:  $\forall u$, the Co-Location Privacy of u is preserved.
\end{definition}

However, even any set of resources in $R$ doesn't violate the co-location privacy of a user $u$, a sly adversary (that know how the principle of the generalization technique works) could try to make some fake resource in order to infer the privacy preference of $u$.
For instance, imagine that Mary (the jealous wife of Bob) suspects a secret meeting between Bob and Alice, for this evening, and she want to enhance her suspect by discovering sensitive privacy settings of Bob in his favorite GeoSN. Mary tries to make a resource $r = \langle ``Mary, Bob, Alice", ``Today, ~22:30p.m.", S, C \rangle$ that means: ``Mary is co-located in C with Alice and Bob". The GeoSN notify Mary that if she want to make this resource she have to set an area that guarantee more distance between users, because she is violating the privacy of someone. With this information, Mary can exclude that she is violating her privacy (she hasn't any privacy setting) and the privacy of any other user outside Alice and Bob, because there isn't any other resource that co-locate other users with Bob or Alice at the time she want to make the resource (She has a complete view on $R$). Thus, Mary can enhance her suspect of a secret meeting between Alice and Bob.

This example shows why we introduced the $Adversary ~Set$ in the privacy settings. If a user $u$ knows who could be a potentially adversary, (s)he can put her/him in the $Adversary ~Set$. The effect is that any resource $r$ that include a user in $\varphi.A$ is allowed even if $r$ violates the privacy preference expressed by $\varphi$.

Note that a sly adversary could exploit her/his inclusion in the $Adversary ~Set$ for a particular kind of attack. For instance, imagine  that Mary (the jealous wife of Bob) suspects a secret meeting between Bob and Alice, and she also suspects her inclusion in the $Adversary ~Set$ of Bob' privacy preferences. If exist some resources that locate Bob somewhere, Mary could makes a fake resource that locate herself near Bob. Thus, any co-location of Bob with Alice (near Mary) is allowed and Mary can enhance her suspects if they occur. Although this kind of attack is possible, we believe that it's difficult to succesfully accomplish, due to the need of multiple necessary conditions for its achievement. Moreover, in contrast to the previous attack, the pubblication of fake resources is required, then the victims can easily became aware of this fact.

\subsection{Co-Location Privacy Preservation}
\label{sec:priv_pres}
In this section, we identify a set of sufficient conditions that $R$ must satisfy in order to guarantee the co-location privacy preservation. These conditions identify a set of possible scenarios that must be avoided in order to have not null the probability of co-locating a user $u$ with an $Excluding ~set ~E$ in a time interval $T$ with a maximum distance greater than $D$.

As mentioned before, the co-location can occur in two ways: a $direct$ way and a $indirect$ way. In the following we formalize these concepts.

\begin{definition}[Direct Co-Location]
A Direct Co-Location of u considering $\varphi \in \Phi(u)$, is a resource r $\in R$:
\begin{equation} \label{eq:direct_loc}
 u \in r.U ~\wedge~ r.U \cap \varphi.E \neq \emptyset ~\wedge~ r.U \cap \varphi.A = \emptyset
\end{equation}
and we indicate it with the syntax: $r \rightarrow (u, \varphi)$.
\end{definition}

A $direct$ co-location must be avoided if it reveals that $u$ and any user in $\varphi.E$ can't be co-located with a distance greater than $\varphi.D$ in the time interval $\varphi.T$. This leads us to define the $validity$ concept for a $direct$ co-location.

\begin{definition}[Valid Direct Co-Location]
A Valid Direct Co-Location of u considering $\varphi \in \Phi(u)$, is a resource r $\in R$:
\begin{equation} \label{eq:valid_direct_loc}
\begin{split}
 r \rightarrow (u, \varphi) ~\wedge~ r.T \notin \varphi.T ~\wedge \\
 \frac{1}{2} \cdot \frac{\varphi.D}{|r.T-n(\varphi.T)|}<V_{max}
 \end{split}
\end{equation}
where n($\varphi.T$) compute the nearest $t \in \varphi.T$ to $r.T$ that is: 
$t \in \varphi.T : |r.T-t|<|r.T-t'| ~ \forall t' \neq t \in \varphi.T$.
\end{definition}

\begin{figure}[direct_loc]
\centering
 \includegraphics[width=1\columnwidth]{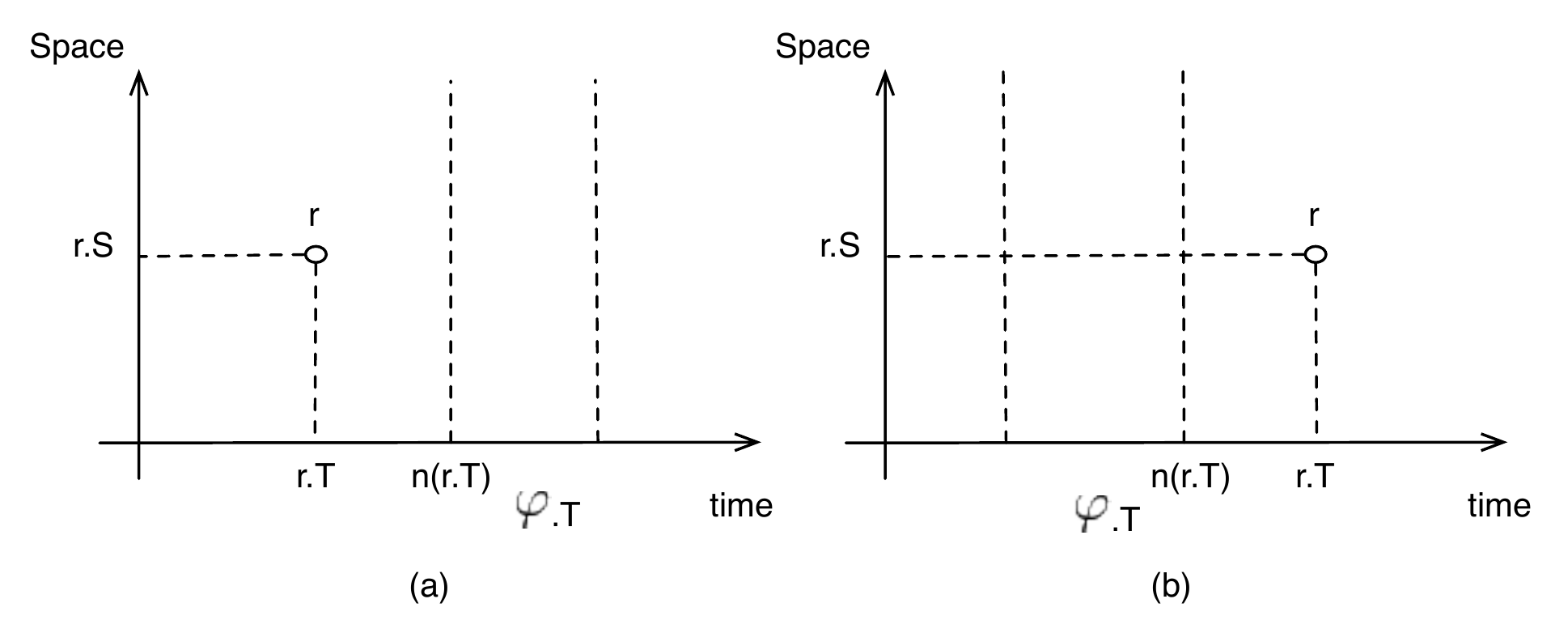}
\caption{Direct Co-Location.} 
\label{fig:direct_loc}
\end{figure}

The last condition condition of (\ref{eq:valid_direct_loc}) states that if a co-location occurs before or after $\varphi.T$ (Fig. \ref{fig:direct_loc}), $u$ and any users in $\varphi.E$ have the possibility of being located with a distance greater than $\varphi.D$ in the time interval $\varphi.T$ because they have enough time to leave each other, if they move with acceptable medium speed. In Fig. \ref{fig:direct_loc}, we report the spatial information of a resource as one-dimensional data. In a real context it would be a two-dimensional data, but this doesn't affect the semantics of our examples.

\begin{definition}[Indirect Co-Location]
An Indirect Co-Location of u considering $\varphi \in \Phi(u)$, is a pair of resource $\langle r,r' \rangle \in R \times R$:
\begin{equation} \label{eq:indirect_loc}
\begin{split}
 r \neq r' ~\wedge~ u \in r.U ~\wedge~ r'.U \cap \varphi.E \neq \emptyset ~\wedge \\
  (r.U \cup r'.U) \cap \varphi.A = \emptyset ~\wedge~  \\
 \frac{\varphi.D-d(r.S-r'.S)}{|r.T-r'.T|}>V_{max}
 \end{split}
\end{equation}
where d(r.S,r'.S) compute the maximum distance between the two location. We indicate it with the syntax: $\langle r,r' \rangle \rightarrow (u, \varphi)$.
\end{definition}

The last condition of (\ref{eq:indirect_loc}) states that the two resources are close in time. This permits us to consider the users involved in the co-location in an area smaller than $\varphi.D$ at the time $max(r.T, r'.T)$.

As a $direct$ co-location, a $indirect$ co-location must be avoided if it reveals that $u$ and any user in $\varphi.E$ can't be co-located with a distance greater than $\varphi.D$ in the time interval $\varphi.T$.

\begin{definition}[Valid Indirect Co-Location]
A Valid Inirect Co-Location of u considering $\varphi \in \Phi(u)$, is a pair of resource $\langle r,r' \rangle \in R \times R$:
\begin{equation} \label{eq:valid_indirect_loc}
\begin{split}
 \langle r,r' \rangle \rightarrow (u, \varphi) ~\wedge~ \\
 [min(r.T,r'.T), max(r.T,r'.T)] \cap \varphi.T \neq \emptyset ~\wedge~ \\
 \frac{1}{2} \cdot \frac{\varphi.D-d(ext(S, |r.T-r'.T|), S')}{|max(r.T,r'.T)-n(max(r.T,r'.T))|}<V_{max}
 \end{split}
\end{equation}
where $S$ is the spatial information associated to the resource with temporal information equals to $min(r.T, r'.T)$, and $S'$ is the spatial information associated to the resource with temporal information equals to $max(r.T, r'.T)$.
\end{definition}

\begin{figure}[indirect_loc]
\centering
 \includegraphics[width=1\columnwidth]{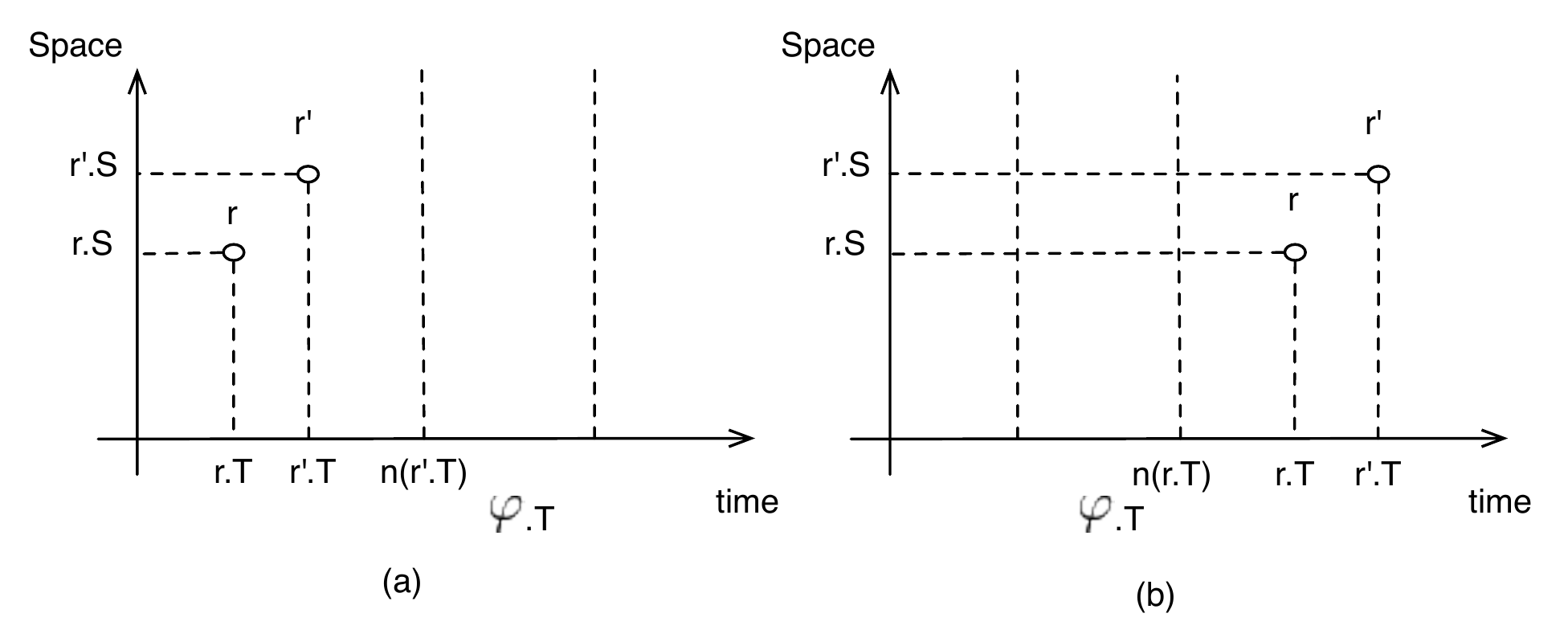}
\caption{Indirect Co-Location.} 
\label{fig:indirect_loc}
\end{figure}

The second condition in (\ref{eq:valid_indirect_loc}) states that the time of the co-location must not overlap with the time interval $\varphi.T$. The last condition states that if the co-location occur before or after $\varphi.T$ (Fig. \ref{fig:indirect_loc}), $u$ and any users in $\varphi.E$ have the possibility (like in the direct co-location) of being located with more distance than $\varphi.D$ in the time interval $\varphi.T$. For example, the indirect co-location in Fig. \ref{fig:indirect_loc} (a) is valid if at the time $r'.T$, $dmax(r.r')<\varphi.D$ and the involved users can depart for at least $\varphi.D$ within $n(\varphi.T)$, moving with an acceptable medium speed.

An important property of a $valid$ co-location ($direct$ or $indirect$) considering $p$, is that if we don't take into account any other resource, the informations given by the co-location are not sufficient to determine that $\forall e \in \varphi.E$ the distance between $u$ and $e$ is not greater than $\varphi.D$ in the time interval $\varphi.T$. This implies that: $\forall t \in \varphi.T ~\forall e \in \varphi.E, ~P(d(u,e)>\varphi.D)>0$.

\begin{figure}[ext2]
\centering
 \includegraphics[width=1 \columnwidth]{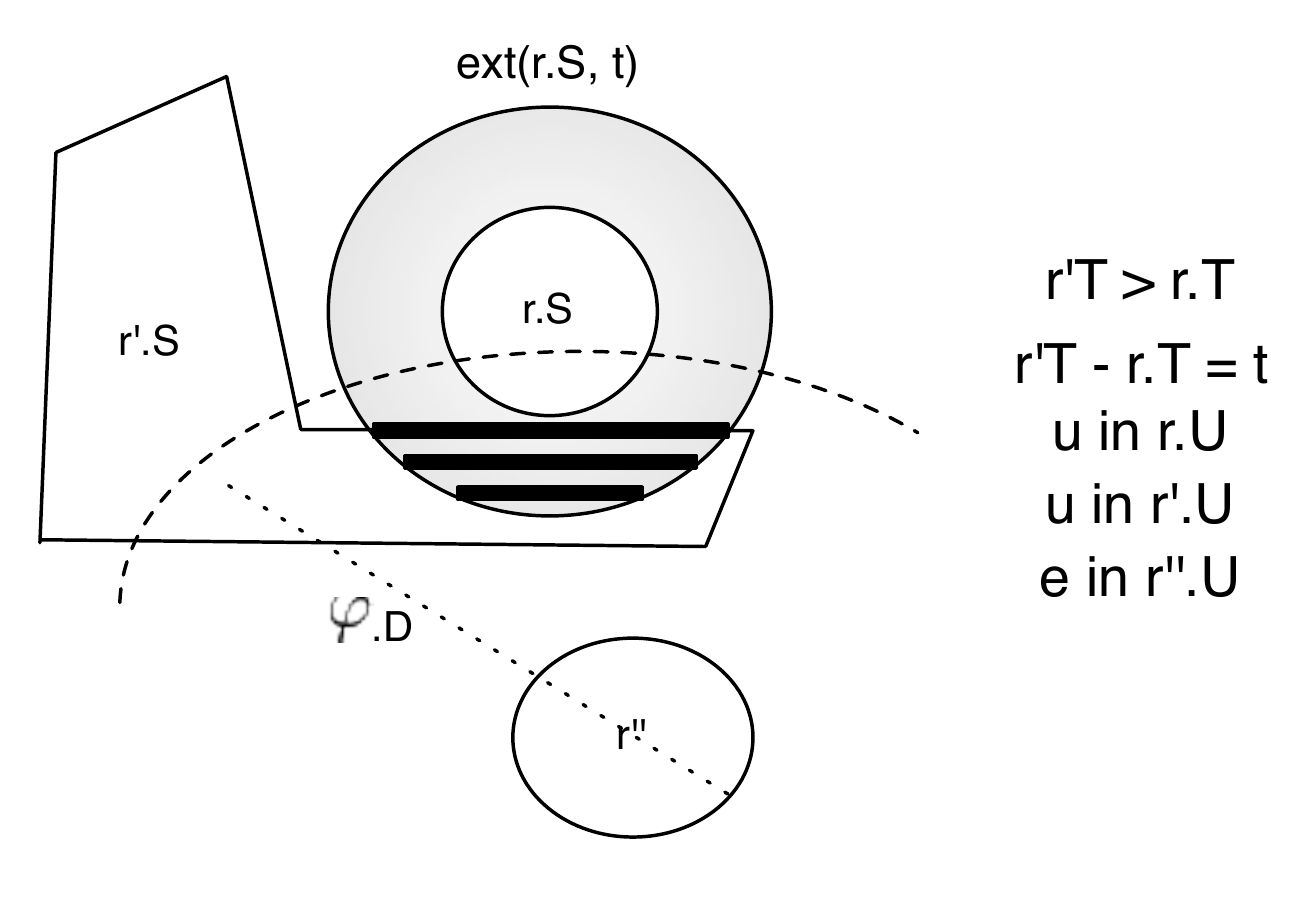}
\caption{Privacy violation in a set of dependent resources.} 
\label{fig:ext2}
\end{figure}

In a set of $dependent$ resources, a $valid$ co-locaiton can still violate the privacy of a user $u$. For instance, let $\varphi=\langle \{e\}, \emptyset, T, D \rangle$ be a a privacy preference for the user $u$. The Fig. \ref{fig:ext2} shows two $valid$ $indirect$ co-location of $u$ considering $\varphi$. But if we consider the gray circle (the area reachable from $r.S$ in the time interval $t$) we can infer that, at the time $r'.T$, $u$ can be located only in the area with black stripes, and this violate the privacy of $u$.

Note that the given example is not valid if we consider a  set of $independent$ resources. This lead us to define two sufficient condition that $R$ must verify in order to preserve the co-location privacy, that are: \\
(a) $\forall \langle r,r' \rangle \in R \times R$, $r \perp r'$ \\
(b) $\forall u ~\forall p \in P(u) ~\forall$ $C$ Co-Location ($direct$ or $indirect$) of $u$ considering $p$, $C$ is $valid$.

Whenever a resource $r$ is added to $R$, it must be $independent$ from any other resource in $R$ and it must not generate any $invalid$ co-locations. If an $invalid$ co-location occurs, we apply a generalization function $g$ defined in section \ref{sec:res}

%
%
%

\section{Co-Location Privacy Preservation Algorithm}
\label{sec:algo}
In this section we propose a sequential algorithm for preserve the co-location privacy in a GeoSN modeled as described in the previous sections. This algorithm takes in input a set of $independent$ resources $R$ that preserve the co-location privacy (the GeoSN resources) and a resource $r$, and it tries to add $r$ to $R$ ensuring that the new set of resources preserve the co-location privacy as well. The algorithm is based on the conditions (a), (b) given in the section \ref{sec:priv_pres}.

Note that in the section \ref{sec:coloc} we state that the set of privacy settings for a user $u$ can be infinite. This problem can be easily avoided by extending the definition of a $privacy$ $preference$ tuple, for example by adding a $frequency$ $flag$ that could assume a value among: \emph{``every day", ``every week", ``every year"}.

Before the execution of the \emph{co-location privacy preservation algorithm}, $r$ is pre-processed to make it $independent$ from any other resource in $R$. This topic is widely discussed in \cite{Freni10}, in the ``WYSE Technique" section. In particular \cite{Freni10} proposes two different algorithms, called $CountryCloakWyse$ and $ClockWyse$ that apply a generalization on the spatial or temporal dimensions, respectively. Since in this paper, we use a $real$-$time$ publication model, only the $CountryCloakWyse$ algorithm is suitable for our purpose.

\begin{algorithm}
\caption{Co-Location Privacy Preservation Algorithm (1)}
\label{alg:priv1}
\begin{algorithmic} 
\Require $R ~set ~of ~independent ~resources, $\\ $r ~resource : R \cup \{r\} ~set ~of ~independent ~resources$
\Ensure $r ~not ~invalid ~direct ~co-location$
\ForAll{$u \in r.U$}
\ForAll{$\varphi \in \Phi(u)$}
\If{$\neg isValidCoLoc(r,u,\varphi)$}
\State $r.U \leftarrow r.U \setminus \varphi.E$
\EndIf
\EndFor
\EndFor
\end{algorithmic}
\end{algorithm}


After a resource is guaranteed to be independent from every other existing resource, the \emph{co-location privacy preservation algorithm} is executed. This algorithm is composed by two different parts. The first part (Alg. \ref{alg:priv1}) is responsible for modifying $r$ if an $invalid$ $direct$ co-location occurs. In this case, only users erasure is applicable.

Instead the second part (Alg. \ref{alg:priv2}) is responsible for avoiding $invalid$ $indirect$ co-locations. The first step of Alg. \ref{alg:priv2} is the computation of the \emph{Co-Location Graph} that is a data structure used to support the \emph{indirect co-location} identification. It is defined as follows:

\begin{definition}[Co-Location Graph]
A Co-Location Graph is a not direct graph G=$\langle V, E \rangle$:
\begin{itemize}
  \item $V$ is the set of vertices and represents a set of resources.
  \item $E \subseteq V \times V$ is the set of edges that link near resources. The edges are enriched with a distance information $d$ (maximum distance between two vertices).
\end{itemize}
\end{definition}

\begin{figure}[graph]
\centering
 \includegraphics[width=0.6\columnwidth]{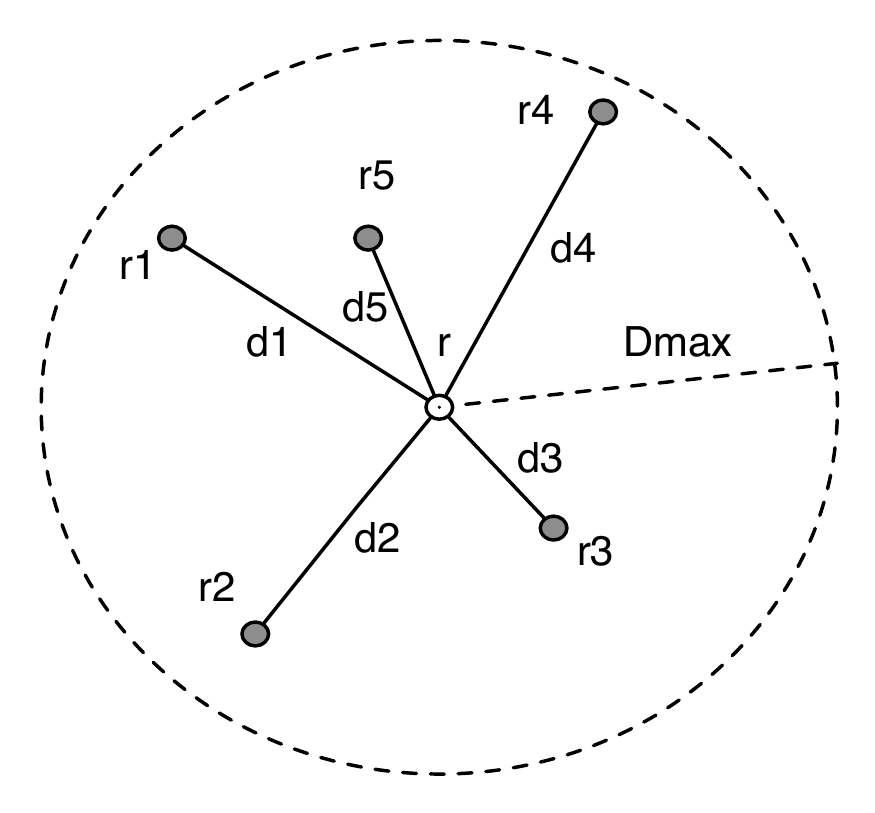}
\caption{Co-Location Graph.} 
\label{fig:graph}
\end{figure}

A \emph{co-location graph} connects any two resources if they are geo-located with a small distance. The \emph{co-location graph} building, in the Alg. \ref{alg:priv2}, starts considering the resource $r$, and $ \forall r' \in R : d(r.S, r'.S)<D_{max}$, the building process creates an edge between $r$ and $r'$ enriched with $d(r.S, r'.S)$. At the end of the process, the data structure $G$ is similar to the graph showed in Fig. \ref{fig:graph}.

\begin{algorithm}
\caption{Co-Location Privacy Preservation Algorithm (2)}
\label{alg:priv2}
\begin{algorithmic}
\Require $R ~set ~of ~independent ~resources, $\\ $r ~resource : R \cup \{r\} ~set ~of ~independent ~resources$
\Ensure $\forall r' \in R, ~\langle r,r' \rangle ~not ~invalid ~indirect ~co-location.$
\State $committed \leftarrow false$
\State $G \leftarrow buildCoLocationGraph(r)$
\ForAll{$\langle r,r' \rangle \in G.E$}
\ForAll{$u \in r.U \cup r'.U$}
\ForAll{$\varphi \in \Phi(u)$}
\If{$\neg isValidCoLoc(\langle r,r' \rangle,u,\varphi)$}
\State $S' \leftarrow enlargement(r.S,r'.S, \varphi.D)$
\If{$S' \neq null$}
\State $r.S \leftarrow r.S + S'$
\Else
\If{$(r.U \setminus u \setminus \varphi.E) \supseteq owner(r)$}
\State $r.U \leftarrow ( r.U \setminus u \setminus \varphi.E)$
\Else
\State $Deny ~r.$
\Return
\EndIf
\EndIf
\EndIf
\EndFor
\EndFor
\EndFor
\Sync
\State $G' \leftarrow buildCoLocationGraph(r)$
\If{$G = G'$}
\State $R \leftarrow R \cup \{r\}$
\State $committed \leftarrow true$
\EndIf
\EndSync
\end{algorithmic}
\end{algorithm}

After the \emph{co-location graph} building is done, the algorithm iterates for all possible $invalid$ $indirect$ co-locations and avoids them by obfuscating the geographical meta-data given by $r$. The function $enlargement(r.S,r'.S, \varphi.D)$ computes an area $S'$ : 

\begin{equation} \label{eq:enlargement}
\begin{split}
 \exists \langle s,s' \rangle \in (r.S+S') \times r'.S : d(s,s') = \varphi .D ~\wedge \\
 \forall r'' \in R,~ r'' \perp \langle r.U, r.T, r.S+S', r.C \rangle 
\end{split}
 \end{equation}

If such an area can't be computed, the algorithm applies a users erasure. If also users erasure can't also be performed, the resource $r$ is denied.
At the end of the Alg. (\ref{alg:priv2}), if there aren't new involved resources in the co-location graph $G$, the resource $r$ is atomically committed to the whole resources set $R$, otherwise ($committed$ = $false$) the algorithm must be re executed. Atomicity is required because multiple instance of the Alg. (\ref{alg:priv2}) can run at the same time due to multiple resource publication from different users at the same time.

After the Alg. (\ref{alg:priv1}) and (\ref{alg:priv2}) are executed, the user is notified about the changes made on $r$ before the pubblication.

\section{Applicability to Existing GeoSNs}
\label{sec:app}
In some GeoSNs, the temporal or the spatial dimension is less crucial and thus it can be generalized if a privacy concern exists. If the resources of a GeoSN doesn't require exact location but require real-time publishing, the technique presented in this paper could be applied in order to preserve the co-location privacy. For instance, Twitter doesn't require that users publish an exact location and they can generalize it to a coarser one (such as a neighborhood or city). In contrast, a tweet's utility generally relies heavily on its publication in real time. Other services require using an exact location, but resources doesn't need to be published instantly. In this case the proposed technique must be extended in order to apply a temporal cloaking instead of a spatial cloaking.
Finally, for services that require both high spatial and temporal accuracy, applying any spatio-temporal cloaking techniques wouldn't be possible. For these services, we can obfuscate the set of users involved in resources by the erasure of some of them. Encryption is also a potentially effective solution.

In the case of a GeoSN allows to omit the geotag, the co-location privacy preservation is more difficult due to an additional kind of attack that the adversary could perform to infer the geo-location of a user. For instance, the adversary may infer the location of a user $u$, if exists a resource $r$ that co-locates $u$ with another user $u'$ without geographical information and another resource $r'$ (with a small temporal difference from $r$) locates $u'$ in a certain place.

In \cite{Vicente11} an overview of the features of existing GeoSNs is given. We can observe that our technique is suitable for services like: Twitter, Google Latitude, Google Buzz, Grindr and Loopt.

\section{Conclusion}
\label{sec:concl}
Since social networking services continue to proliferate, there is an increasing need of preserving users privacy. This paper addresses a particular privacy threat that is the \emph{co-location privacy}.
We propose a way for expressing privacy settings of users and a study of how they can be preserved in the context of ``real-time publishing" GeoSNs. The paper formalizes the setting, provides a way for easily defining privacy preferences, and provides a technique that generalizes the tags of resources so that these remain useful while ensuring that the privacy preferences are preserved. This technique exploit spatial generalization or users erasure (where the first one is not possible). In this paper we take into account only meta-data of resources to preserve users privacy, but the content of a resource itself could raises privacy threats. For example a photo can co-locates some users against their will. Future research can be done in this direction to face this problem. Moreover, the proposed technique can be extended in order to face the co-location privacy threat in GeoSNs that don't require real-time publishing. In this case, temporal cloaking of resources is also possible.

\balance

\bibliographystyle{plain}
\bibliography{report}

\end{document}